\documentclass[fleqn,12pt]{SelfArx} 

\definecolor{color1}{RGB}{0,0,185} 
\definecolor{color2}{RGB}{0,20,20} 

\usepackage{hyperref} 
\hypersetup{hidelinks,colorlinks,breaklinks=true,urlcolor=color2,citecolor=color1,linkcolor=color1,bookmarksopen=false,pdftitle={Title},pdfauthor={Author}}
\JournalInfo{Accepted for publication by Astronomy Reports, 2018 No 1}

\PaperTitle{
Can superflares occur on the Sun?\\
A view from dynamo theory
} 

\Authors{M.\,M.~Katsova$\,$\textsuperscript{1}*, ~ 
 L.\,L.~Kitchatinov$\:$\textsuperscript{2,3}, ~ 
 M.\,A.~Livshits$\,$\textsuperscript{4},\\
 D.\,L.~Moss$\,$\textsuperscript{5}, ~ 
 D.\,D.~Sokoloff$\;$\textsuperscript{6}, and ~ 
 I.\,G.~Usoskin$\,$\textsuperscript{7}} 
 
\affiliation{\textsuperscript{1}\textit{
Sternberg State Astronomical Institute, Lomonosov Moscow State University, 119991, Moscow, Russia
\href{mailto:maria@sai.msu.ru}{\it ~~ maria@sai.msu.ru}
}} 
\affiliation{\textsuperscript{2}\textit{
Institute for Solar-Terrestrial Physics, PO Box 291, Irkutsk, 664033, Russia  \href{mailto:kit@iszf.irk.ru}{\it ~~ kit@iszf.irk.ru}
}} 
\affiliation{\textsuperscript{3}\textit{
Pulkovo Astronomical Observatory, St. Petersburg 196140, Russia
}} 
\affiliation{\textsuperscript{4}\textit{
IZMIRAN, 4 Kaluzhskoe  Shosse, Troitsk, Moscow, 108840, Russia \href{mailto:maliv@mail.ru}{\it ~~ maliv@mail.ru}
}} 
\affiliation{\textsuperscript{5}\textit{
School of Mathematics, University of Manchester, Oxford Road, Manchester, M13 9PL, UK  \href{mailto:david.moss@manchester.ac.uk}{\it ~~ david.moss@manchester.ac.uk}
}} 
\affiliation{\textsuperscript{6}\textit{
Department of Physics, Moscow University, 119992 Moscow, Russia \href{mailto:sokoloff.dd@gmail.com}{\it ~~ sokoloff.dd@gmail.com}
}} 
\affiliation{\textsuperscript{7}\textit{
Space Climate Research Unit and Sodankyl\"a Geophysical Observatory, 90014 University of Oulu, Finland \href{mailto:ilya.usoskin@oulu.fi}{\it ~~ ilya.usoskin@oulu.fi}
}} 

\newcommand{\keywordname}{Keywords}

\Abstract{
\small Recent data from the {\sl Kepler} mission has revealed the occurrence of superflares in sun-like stars 
which exceed by far any observed solar flares in release of energy. Radionuclides data do not provide evidences 
for occurrence of superflares on the Sun over  the past eleven millennia.  Stellar data for a subgroup of 
superflaring {\sl Kepler} stars are analysed in an attempt to find possible progenitors of their abnormal 
magnetic activity. A natural idea is that the  dynamo mechanism in superflaring stars differs in some respect 
from that in the Sun. We search for a difference in the dynamo-related parameters between superflaring stars and 
the Sun to suggest a dynamo-mechanism as close as possible to the conventional solar/stellar dynamo but capable 
of providing much higher magnetic energy. Dynamo based on joint action of differential rotation and mirror 
asymmetric motions can in principle result in excitation of two types of magnetic fields. First of all, it is 
well-known in solar physics dynamo waves. The point is that another magnetic configuration with initial growth 
and further stabilisation is also possible for excitation. For comparable conditions, magnetic field strength of 
second configuration is much larger rather of the first one just because dynamo do not spend its efforts for 
periodic magnetic field inversions but use its for magnetic field growth. We analysed available data from the 
{\sl Kepler} mission concerning the superflaring stars in order to find tracers of anomalous magnetic activity.
Starting from the recent paper \cite{1}), we find that anti-solar differential rotation 
or anti-solar sign of the mirror-asymmetry of stellar convection can provide the desired strong magnetic field in 
dynamo models. We confirm this concept by numerical models of stellar dynamos with corresponding governing 
parameters.  
We conclude that the proposed mechanism can plausibly explain the superflaring events at least for some cool 
stars, including binaries, subgiants and, possibly, low-mass stars and young rapid rotators.
}

\begin{document}

\flushbottom 

{\parindent0pt
\maketitle 
}
\thispagestyle{empty} 
\pagebreak

\tableofcontents 


\section{Introduction}

Recent data from the {\sl Kepler} mission \cite{2} has revealed the existence  of stellar "superflares" \cite{3,4}.
Some of the superflaring stars are  G-dwarfs
with long rotation periods $P_{\rm rot} > 10$ days \cite{4}, and two of these stars are very similar to the Sun in 
their surface temperature  and rotation rate \cite{5}
Observational techniques reveal flares with total energy
substantially greater than $10^{33}$ erg (to be compared with the highest energy,
approximately $10^{32}$ erg, of any  observed solar flares -- see details in Sect.\ref{strong})
however sometimes the reported energy is as large as $10^{36}$ erg.

Observations of superflares on Sun-like stars are challenging for solar physics in general and for the solar 
dynamo in particular.
On one hand, the energy of the solar magnetic field is insufficient to produce a superflare (assuming a 
reasonable  efficiency of the flare production \cite{6}).
On the other hand, cosmogenic isotope studies have identified historical events that were 30-50 times stronger 
(in the sense of the fluxes of solar energetic particles) than the most energetic solar flares observed 
instrumentally \cite{3,7,8}. It is likely that these events form the upper limit of the intensity of solar events 
during the last eleven millennia (see Sect.\ref{strong}), while the strongest stellar superflares are 
substantially stronger.

An important question remains whether superflares with energies up to $10^{35}$\,--\,$10^{36}$ erg are possible 
on the present-day Sun. If not, what is the physical difference between the Sun and the superflaring Sun-like 
stars? We stress that this question is beyond purely academic interest because a solar superflare would be 
hazardous for modern technology.
\cite{9} noticed that the dependence of superflare occurrence frequency on the flare energy found in \cite{3} for Kepler 
targets forms a high-energy continuation of the power-law distribution of the observed solar flares. From this 
interpolation they suggested an occurrence frequency of flares on the Sun to be one flare with energy 
$\geq10^{34}$\,erg per 800 years and one flare with energy $\geq10^{35}$\,erg per 5000 years.
However, this disagrees with the radionuclide data for the last $\sim 10^4$ years \cite{10,11} 
(see also  Section~\ref{strong}) and data from lunar rocks \cite{12}).

\section{A possible origin of superflares}

A step towards addressing the problem was recently undertaken by Kitchatinov \& Olemskoy \cite{1} who suggested a 
dynamo model based on fluctuations of the dynamo drivers. The fluctuations can drive a solar activity cycle with 
sufficient magnetic energy to produce flares stronger than flares observed astronomically and similar to those 
identified in isotopic data. By varying the model parameters even stronger flares can possibly  be explained, but 
hardly the strongest observed stellar flares of the energy of ca. $10^{36}$ erg.

The key idea of \cite{1} is as follows. The intensity of dynamo drivers can be accumulated 
in one dimensionless dynamo number $D$
\begin{equation}
    D = \frac{\Delta\Omega\alpha R^3_\odot}{\eta^2_{_\mathrm{T}}},
    \label{1}
\end{equation}
which combines the differential rotation ($\Delta\Omega$), the eddy diffusivity ($\eta_{_\mathrm{T}}$) and the 
rate of toroidal-to-poloidal field conversion by cyclonic motions ($\alpha$) into a single parameter quantifying 
their common efficiency in producing magnetic activity. With a conventional definition, $D$ is positive for 
solar-type differential rotation with positive $\alpha$ (in the Northern hemisphere). 
Note, that reversal of the sign of dynamo number requires either reversal of the sign of $\alpha$ {\it or} the 
sign of $\Delta \Omega$.
A reversal of the sign of dynamo number, to become negative, switches the dynamo into a regime with much higher 
magnetic energy. The expected fluctuations in the dynamo drivers can cause the dynamo number to fluctuate. 
Sufficiently large fluctuations can produce rare events of sign reversals of $D$ yielding substantially higher 
magnetic energy and leading, subsequently, to more energetic flares.

The present paper develops this idea further and suggests that Sun-like stars with extremely energetic flares can 
in fact be different from the Sun. This difference results in dynamo numbers which are regularly negative (rather 
than positive as in the solar case).
The statistics of superflares on Kepler stars does not exclude this possibility. It is found \cite{4,13}  that only 
about 0.3\%~ of Kepler targets display superflares. 
Moreover, stars in this small group exhibit a very uneven distribution of flares with $E > 10^{34}$ erg, having on 
average about three superflares per a star per year. 
Stars of this small group may differ from the bulk of the population of less active stars by hosting dynamos with 
negative $D$.

Let us present the idea of this paper in a more extended form. Dynamo based on joint action of differential 
rotation and mirror asymmetric convection (or turbulence) can in principle generate two types of magnetic 
configuration. First of all, that are conventional for solar physics waves of magnetic field (dynamo waves). The 
point is another configuration with initial magnetic field growth and then saturation is also possible. This 
second type of configurations  occurs in particular for galactic magnetic fields which do not demonstrate traces 
of wave-like behaviour. For comparable conditions, magnetic field strength og magnetic fields of the second type 
is much larger rather for the first one just because abilities of dynamo are not spending on periodic magnetic 
field reversals rather for magnetic field growth only.   Of course, the type of magnetic configurations excited 
in a particular celestial body depends on the flow structure and geometry of the body however sperical dynamos do 
demonstrate a region in parametric space with excitation conditions for both types of magnetic fields are close 
enough to allow excitation of a steady magnetic field instead of an oscillating one. 

Paper \cite{1} supposes that a transition from an oscillating to the stationary configuration appears for a limited 
time as a fluctuation. Here we attract attention to the fact that for many stars (say, for binaries) 
hydrodynamics may be rather different from the solar one and we can expect generation of stationary magnetic 
fields. In this case, magnetic field strength can be much larger than the solar one what in turn  can results in 
superflares. 

We demonstrate that a dynamo model with negative dynamo number can produce magnetic energy which is several 
orders of magnitude higher than a similar model with positive dynamo number. We discuss possible physical reasons 
for stars to have  negative dynamo numbers.

An observational and theoretical understanding for the differential rotation of superflaring stars is obviously 
crucially important for dynamo modelling. Differential rotation of stars observed by Kepler is discussed in \cite{14}.

\section{The strongest solar flares and related events}
\label{strong}

Solar flares have been known since 1859 when a giant white flare was discovered by R.~Carrington and R.~Hodgson, 
and were well studied during the 20th and 21st centuries. However, no records of optical flares are known before 
that. On the other hand, extreme solar events can be studied even for earlier times, using indirect proxy data.
Such a proxy, recoverable with stable resolution for about ten millennia backward, is the cosmogenic radionuclide 
record preserved in a natural terrestrial archive.
The most common and useful cosmogenic proxies are $^{14}$C (radiocarbon), stored
and measured in dendrochronologically dated tree trunks, and $^{10}$Be, measured in polar ice cores \cite{15}.
These cosmogenic proxies are used to reconstruct the flux of galactic cosmic rays (GCR) in the past \cite{16}.
Since the GCR flux is modulated by solar/heliospheric magnetic fields, its variability reflects the (inverse) 
solar activity.
While GCR are always present in the vicinity of the  Earth and are subject to solar modulation, sporadic events 
of solar energetic particles (SEPs) can occur sometimes.
Such events are related to solar eruptive processes, solar flares or coronal mass ejections, which can accelerate 
solar particles (mostly protons) to high energies, up to several GeV, which is sufficient to initiate a nucleonic 
cascade in the Earth's atmosphere and thus to produce cosmogenic nuclides.
However, because of the relatively low time-resolution (annual at best) and high level of noise in the data,
a SEP event must have high magnitude (an "extreme SEP event") in order to be observable in the proxy record \cite{17}.

Presently, only two such extreme SEP events have been discovered in the past.
The greatest event around 775 AD was found in high-resolution $^{14}$C tree-ring data from Japan \cite{18} and low
resolution $^{14}$C and $^{10}$Be data \cite{19}, and later confirmed with other high-resolution $^{14}$C \cite{7,20,21}
and $^{10}$Be and $^{36}$Cl records \cite{22}.
Although various hypotheses were proposed to explain the event, its solar origin has become a paradigm \cite{7,8}.
The strength of the 775 AD event was estimated to be  about $ 40$ times greater than the strongest SEP event of 
the instrumental era (23 Feb, 1956).
Another similar event was found in the high-resolution $^{14}$C data corresponding to the year 994 AD
\cite{23}, and it is estimated to be a factor 1.5 -- 2 weaker than that of 775 AD.
Interestingly, the Carrington event, which produced a strong white-light flare and the strongest
recorded geomagnetic storm \cite{24}  was not accompanied by a strong SEP event \cite{17,19}.

The SEP event of 775 AD may conservatively serve as the strongest event over the Holocene
 (the current interglacial period that started about 11 millennia ago).
Although only a small fraction of this period is covered by high-resolution annual $^{14}$C data, being mostly
 measured with 5-year resolved
 (low-resolution) data, it is unlikely that an event stronger than that took place during the Holocene.
The team led of the paper \cite{25}, which discovered both 775 AD and 994 AD events, has recently performed a 
systematic search for other events,
 starting from the high-resolution $\Delta ^{14}$C measurements around several sharp peaks distinguishable in the 
 low-resolution
 (decadal average) INTCAL dataset.
An analysis of the strongest peak about ($0.4$ {permil/year}) around 775 AD led to the discovery of the event.
This peak was so strong that it was clearly observable even in the low-resolution INTCAL record \cite{19}.
Another smaller peak, hardly distinguishable in the low-resolution INTCAL record, corresponded to the discovery 
of a weaker event in 994 AD.
However, an analysis of high resolution $\Delta ^{14}$C measurements around several other peaks ($\approx 0.3$ 
{permil/year})
 over the last several millennia did not result in finding of new SEP events.
Nevertheless there are several relatively strong peaks ($0.3 \div 0.4$ {permil/year}) in the
low resolution INTCAL data over the Holocene, which are not yet covered by high-resolution data.
These may potentially contain signals of SEP-like sharp events, but they cannot be much stronger than the event 
of 775 AD, otherwise
they would have been detected in a systematic search over the low-resolution INTCAL dataset \cite{26}.

Thus, although we cannot at the moment say that the event of 775 AD was the only of a kind, it can be 
conservatively stated that
 an event stronger than that did not occur during the Holocene.
Accordingly, the event of 775 AD can be considered as the worst case scenario over the last eleven millennia.

We note that the relation between strong flares and extreme SEP events is not one to one.
Not every strong flare would produce an SEP event at Earth, depending on the flare's location on the solar disc 
(flares near the west
 limb are more effective).
This is clearly seen on the example of the Carrington flare which was not accompanied by a strong SEP event.
On the other hand, an extreme SEP event should have a parent flare and a strong coronal mass ejection.
For example, the greatest directly observed event of 23 Feb, 1956 had a strong parent flare of class at least 3-B 
(it occurred partly behind
 the west limb and accordingly its intensity might have been underestimated).

Several of the largest solar flares have been observed in the current epoch, such as events in February 1956, 
August 1972, 1989, October 2003
 with total energies close to those of the Carrington flare. 
Very high-quality data on the total vector of the magnetic field in active regions and the large-scale magnetic 
field including the dipole field of the Sun as a star
 are available now. 
These observations make it possible to calculate the free energy of the magnetic field in active regions that can 
be released in flares. 
These estimates show that even the largest active regions on the Sun are capable of producing non-stationary 
processes (flares and CME) with
 total energy not greater than $3 \times 10^{32}$ erg. 
Such an upper limit for a given active region follows  from energy considerations, namely from the magnetic 
virial theorem as well \cite{27}.
As for the magnetic fields on active F, G, and K main-sequence stars, there are modern spectropolarimetric
observations indicating that some fast rotating, young sun-like G stars possess magnetic fields around 5 G \cite{28}. 
Spots on these stars can cover up to 10\%~ of a stellar surface. 
These results, together with the magnetic virial theorem, imply that the maximal
possible energy of the strongest flare on these stars cannot exceed  $10^{34}$ erg \cite{29}.
Stronger events require for their explanation, either another origin of flares or changes to the dynamo 
mechanism.

\section{Dynamo modelling}

\begin{figure}
\includegraphics[width=0.8\textwidth]{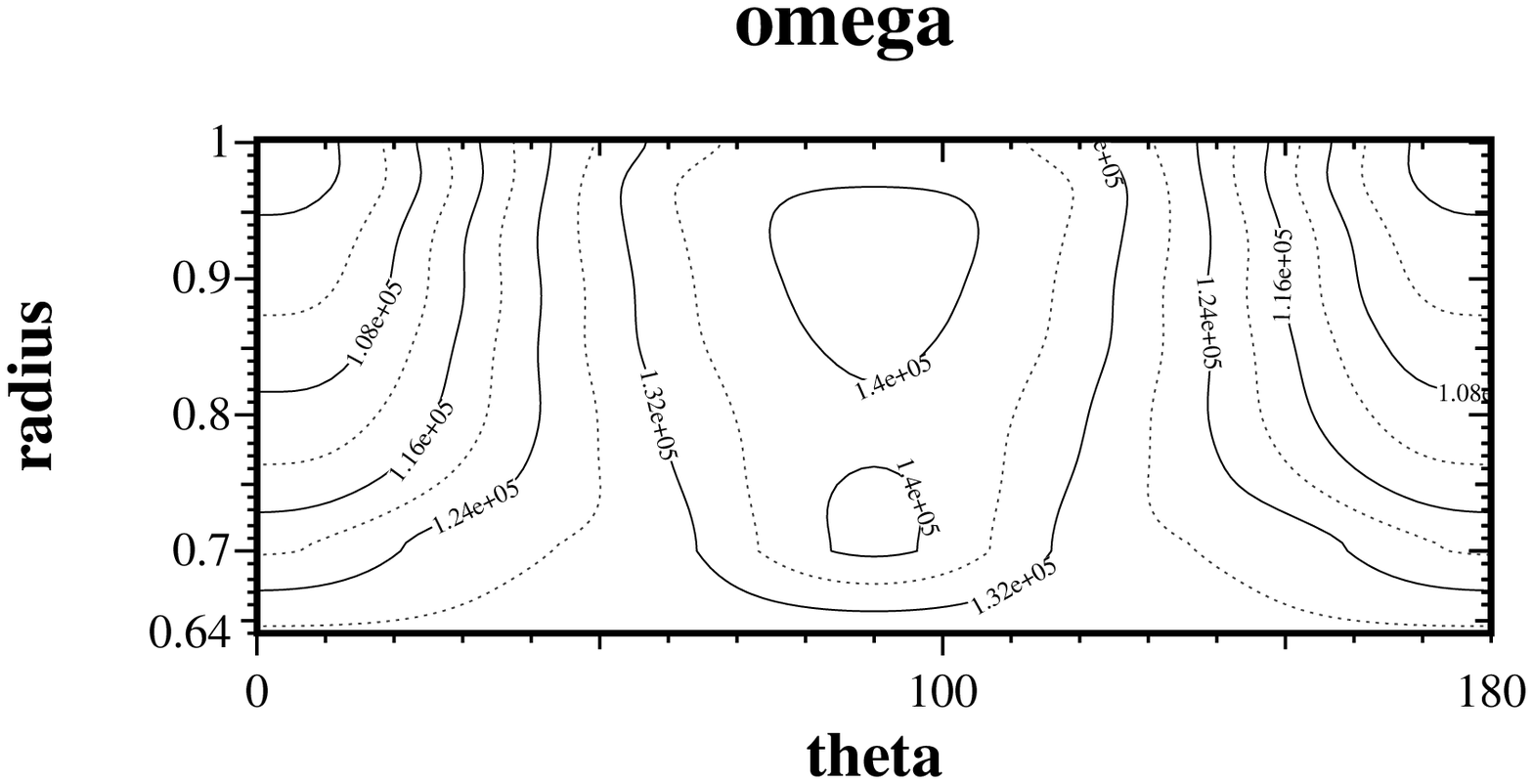}\\[20pt]
\includegraphics[width=0.8\textwidth]{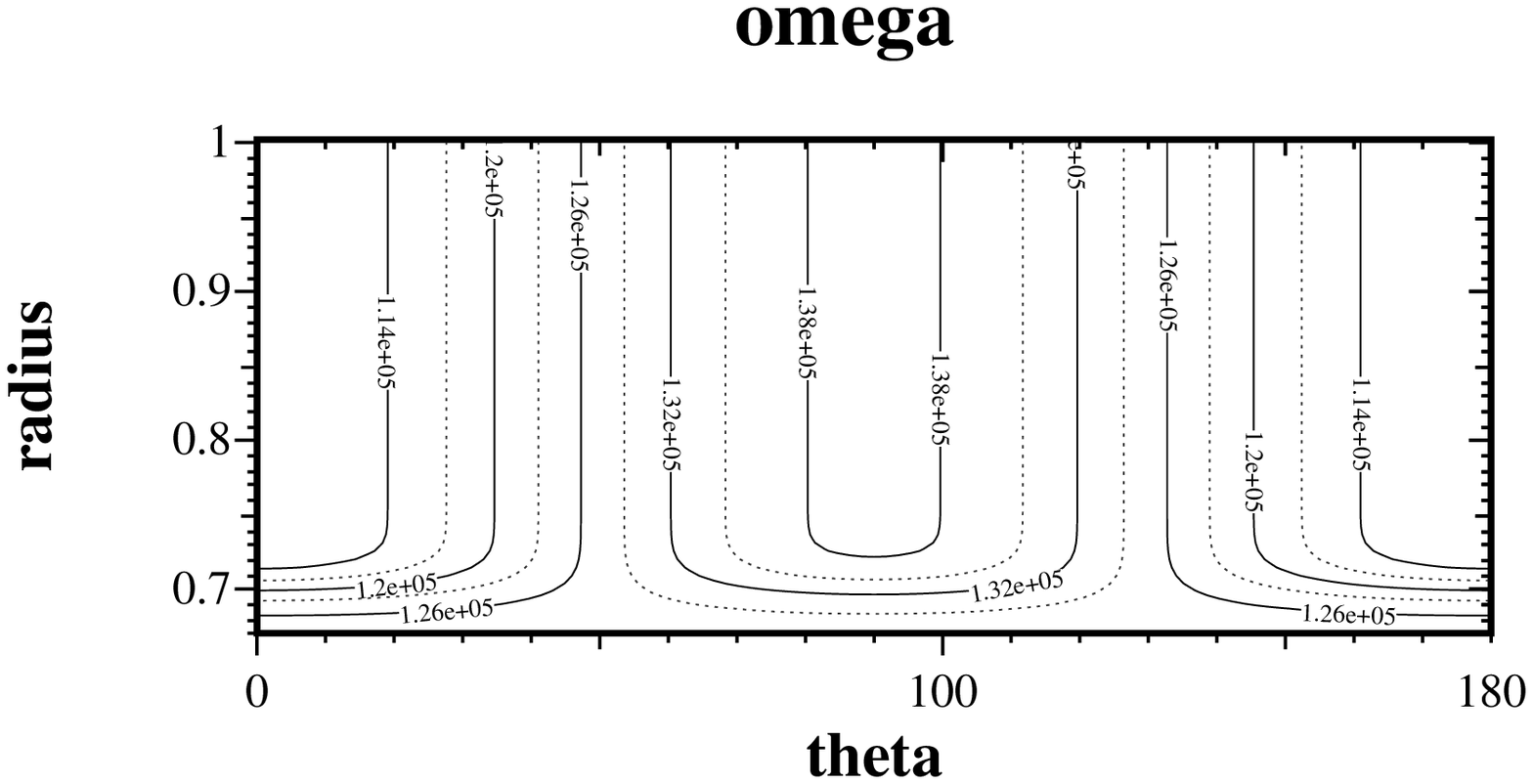}\\[10pt]
\caption{\normalsize\sl Stellar rotation curves: upper panel -- SOHO-like rotation curve, 
lower panel -- \cite{32} 
rotation curve.}
\label{rotcurve}
\end{figure}

In order to demonstrate the dependence of the dynamo-generated magnetic energy on the sign of the dynamo number, 
we use a conventional mean-field model for dynamo in spherical geometry driven by the joint action of the 
differential rotation and a mirror asymmetric factor $\alpha$ responsible for restoration of the poloidal 
magnetic field from toroidal (e.g. \cite{30}). The physical nature of the factor $\alpha$ can be associated with 
the action of magnetic or Coriolis forces. Which of them is more important for stellar dynamos is not crucial 
for what follows. For the sake of definiteness and simplicity we use the parametrization

\begin{equation}
\alpha = \frac{\alpha_0(r)}{1+\vec{B}^2({\vec r}, t)}
\label{alphpar}
\end{equation}
This is  the simplest nonlinear suppression of dynamo action, known as algebraic
$\alpha$-quenching. In principle, it would not be problematic to include various additional effects, such as 
meridional circulation and/or  more sophisticated types of dynamo quenching. For the sake of definiteness 
we use two different parametrizations of the stellar rotation curve (Fig.\,\ref{rotcurve}), available in literature, 
viz. a SOHO-like rotation law (e.g. \cite{31}) and the rotation law \cite{32}.

The key point of our idea is that the dynamo drivers are not very well-defined quantities and there is a 
probability that the sign of the main dimensionless number, which determines stellar dynamo action in a Sun-like 
star, can be opposite to that inferred for the Sun.
Physical reasons for such an idea are discussed in the following section while here we consider some consequences 
of this hypothesis.

We run dynamo models with both rotation curves of Fig.\,\ref{rotcurve} and then reverse the sign of $D$ and see 
what happens with magnetic energy after a steady state is reached.
With the SOHO-like rotation curve (Fig.\,\ref{rotcurve}, upper panel), the result is quite straightforward. The 
sign of $D$ determines the direction of dynamo wave propagation so  that the activity wave reverses its direction 
of propagation from equatorward to poleward while the magnetic energy remains more or less the same as before. We 
do not illustrate this rather standard option in a figure.

\begin{figure}
\includegraphics[width=0.8\textwidth]{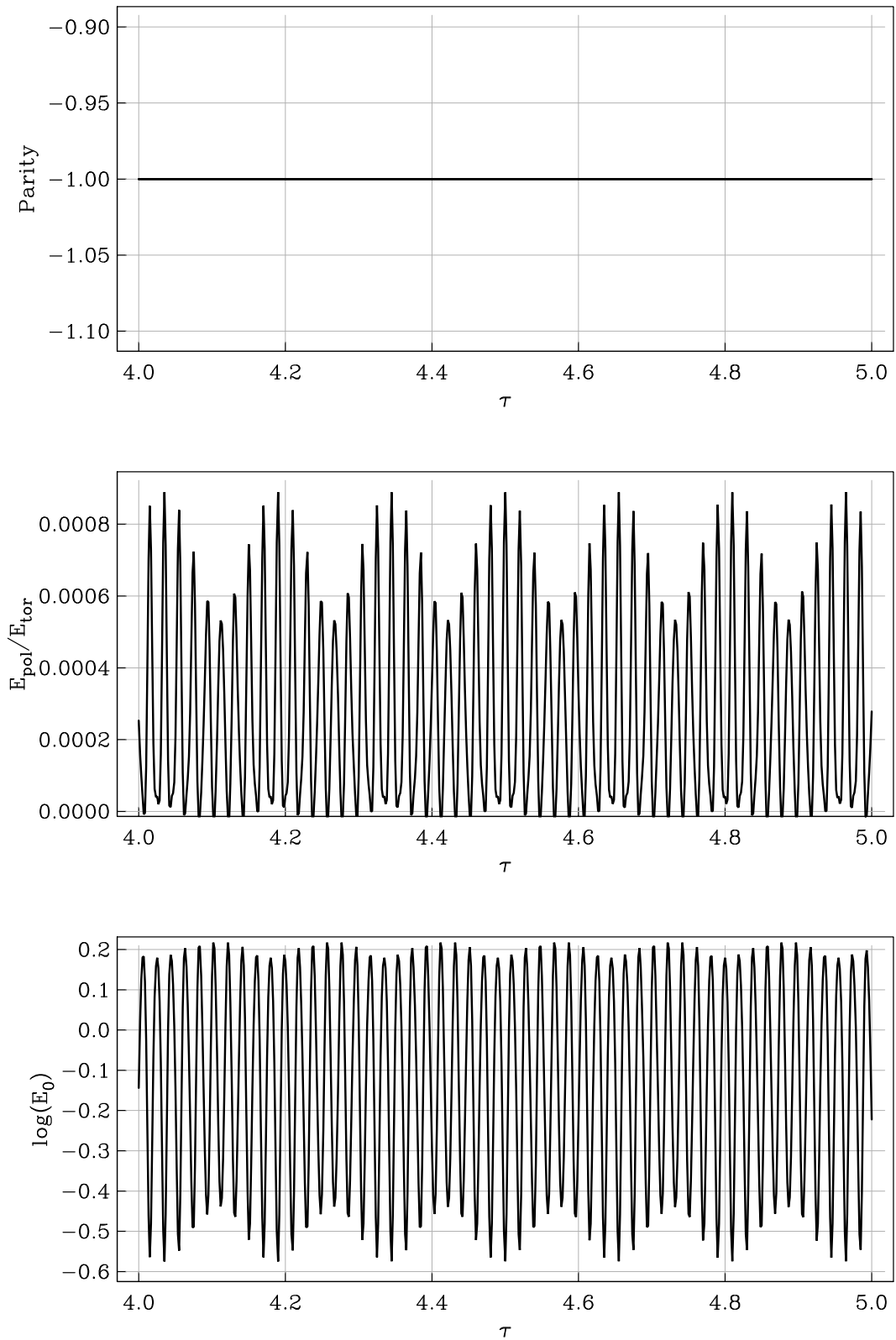}
\caption{\normalsize\sl Magnetic field for  rotation law \cite{32}, $D>0$, timeseries for parity (top), ratio of
magnetic energies of toroidal and poloidal magnetic fields (middle) and total energy of the mean magnetic
field  (bottom).}
\label{Dpos}
\end{figure}

The \cite{32} rotation law gives a much more instructive result.
Time series for the standard solar case $D>0$ are presented in Fig.\,\ref{Dpos}.
We conclude that we obtain a standard travelling wave of solar type.
A deeper investigation of the result shows that the main direction of dynamo wave propagation is radial, while 
some projection of the wave vector on the meridional direction provides an equator-ward propagation of activity wave (cf. \cite{33}).
For this case the amplitude of $\alpha$ is chosen to be about twice the  marginal value for dynamo action.

\begin{figure}
\begin{center}

\includegraphics[width=0.8\textwidth]{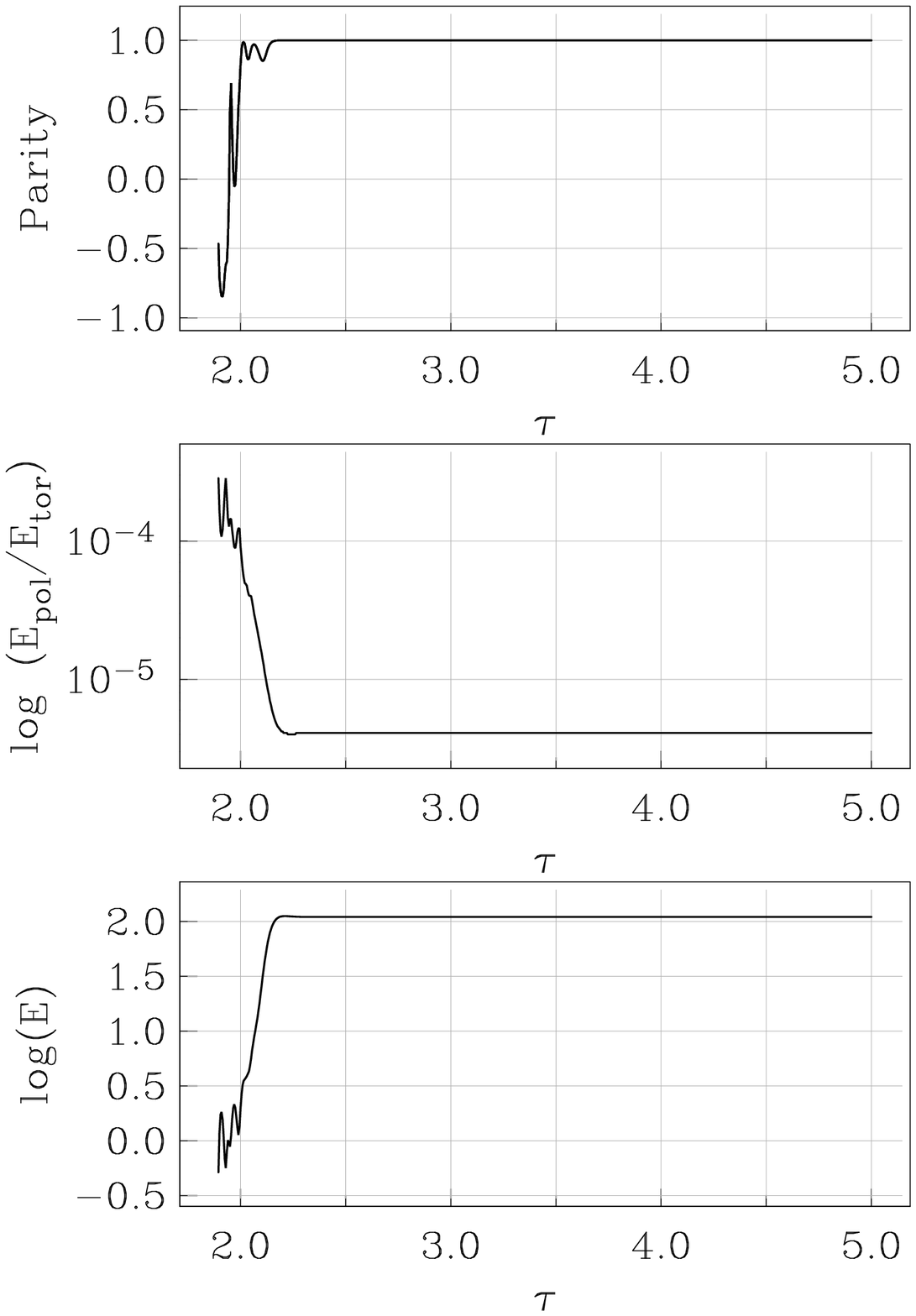}

\end{center}
\caption{\normalsize\sl Summary of the fields obtained using the  rotation law \cite{32}, $D<0$, 
time series for parity (top), ratio of
magnetic energies of toroidal and poloidal magnetic fields (middle) and total energy of the mean magnetic
field  (bottom). The asymptotic ratio of $E_{\rm pol}/E_{\rm tor}$ is approximately $4.1 \times  10^{-6}$.}
\label{Dneg}
\end{figure}

Reversing the sign of the dynamo number $D$ (Fig.\,\ref{Dneg}) we obtain a steady dynamo producing a magnetic 
configuration with substantially higher magnetic energy
compared with the previous case ($\log E \approx 2.0$ vs. $\log E \approx 0.2$, respectively).

One more instructive point is that the steady ratio between toroidal and poloidal magnetic fields for the 
magnetic configuration with sign reversed $D$ is about one order of magnitude lower then the average value of 
this ratio for the oscillating configuration ($E_{\rm pol}/E_{\rm tor} \approx 4.1 \times  10^{-6}$ to be 
compared with
$6 \times 10^{-4}$  in Figure 2, taking into account that here we deal with quadratic quantities). It means that 
the strong magnetic field is now more hidden in the stellar interior and it is less observable between the 
superflares.

\section{From the classification of superflaring stars to the sign of dynamo number}

\begin{table*}
\caption[]{\normalsize\sl Some solar-type stars with superflares with $E> 10^{35}$ erg, after \cite{34}. 
$T_{\rm eff}$ is the effective temperature, $g$ is the gravity in cm\,s$^{-2}$, $P_{\rm rot}$ 
is the rotation period.}
\begin{center}
{\small
\begin{tabular}{|c|c|c|c|c|c|c|}
\hline
KIC number & $T_{\rm eff}$ K , & $\log g$ & $\log E$ &   N & $P_{\rm rot}$, day & Comments\cr
\hline
\multicolumn{7}{|c|}{\bf Binaries}\cr
\hline
1156431 & 5094 & 4.514 & 35.06 & 213 & 3.142 & Binary/multiple system \cite{35}, unusual differential  
\cr
        &      &       &       &     &       & rotation \cite{36}, cooler than the Sun \cr
\hline
8481574 & 5722 & 4.487 & 35.06 & 2 & 0.326 & eclipsing binary$^1$  \cr
\hline
9752973 & 5865 & 4.017 & 35.03 & 1 & 13.05 & eclipsing binary$^1$ \cr
\hline
12156549 & 5541 & 4.378 & 36.50 & 128 & 3.651 & binary$^1$; oscillations in superflares \cite{37} \cr
\hline
9655129 & 5140 & 4.431 & 35.38 & 26 & unknown & detatched Algole-type$^1$ \cr
\hline
\multicolumn{7}{|c|}{\bf Subgiants} \cr
\hline
6437385 & 5401 & 3.713 & 36.78 & 18 & 13.672 & oscillations in superflares \cite{37} \cr
\hline
7350496 & 5453 & 3.744 & 36.43 & 4 & 9.403 &  \cr
\hline
8226464 & 5754 & 4.053 & 36.44 & 15 & 3.101 & quasi-periodic pulsations \cite{38} \cr
\hline
\multicolumn{7}{|c|}{\bf Stars cooler than the Sun}\cr
\hline
3945784 & 4854 & 4.444 & 35.09 & 18 & 15.267 & K2 V \cr

\hline
\multicolumn{7}{|c|}{\bf Stars with pulsations or oscillations}\cr
\hline
5475645 & 5336 & 4.654 & 35.63 & 6 & 7.452   & quasi-poeriodic pulsations \cite{37}\cr
\hline
11610797 & 5865 & 4.465 & 35.76 & 34 & 1.625 &  oscillations in superflares \cite{37} \cr
\hline
\multicolumn{7}{|c|}{\bf Very young fast rotating stars} \cr
\hline
9652680 & 5618 & 4.802 & 35.38 & 26 & 1.408 &  \cr
\hline
\end{tabular}
}
\end{center}
\footnotesize\it
$^1$Information concerning stellar variability is added from the SIMBAD database, provided by CDS, Strasbourg.
\end{table*}

We expect that hydrodynamics of a particular star with superflares differs somehow from solar hydrodynamics in 
order to provide a negative dynamo number $D$ what in turn gives a much  stronger magnetic field compared to the 
Sun. Developing a hydrodynamical model for such stars
needs to be addressed specifically and it is obviously beyond the scope of this paper.
Here we limit ourselves to presenting a list of solar-type stars with superflares from \cite{34} which gives a hint 
that its hydrodynamic may be substantially different from in the Sun.

Indeed, observations of stellar superflares by the Kepler mission show that energies of most events do not exceed 
$10^{34}$ erg. 
However, among sun-like stars with superflares that are characterized in \cite{34} as rotationally variable, 
there are some objects where 
 more energetic superflares do occur. 
We try to check whether these stars are solar-type or different. 
We choose stars with superflares with energies $> 10^{35}$ erg, which comprise
 more than 15\%~ of the list in \cite{34}. 
Several examples of such stars are presented in Table 1. 
An analysis of the properties of the selected stars indicates that the pattern of their variability is 
significantly different
 from solar-type activity (see comments to Table 1).

These examples are  illustrative only and do not  pretend to represent a full choice of possibilities.
Additional options include binaries such as the Algol variables (e.g. \cite{39}). Standard models of stellar 
hydrodynamics used in dynamo studies assume that stars can be considered as single objects (see, however, 
\cite{40,41}). Tidal interactions in stellar convective shells may be associated with non-solar distribution of dynamo 
drivers. Another possibility here is related to subgiants.

The sign of the dynamo number (see definition of $D$ in Eq.~(\ref{1})  can be opposite to the solar case if 
either the differential rotation is anti-solar or the sign of $\alpha$ is reversed. Strassmeier \cite{42}  reviewed 
similarities and differences in magnetic activity between cool stars and the Sun. His Table\,1 gives several 
examples of observed anti-solar rotation. All of them belong to close binaries or giants. For MHD modelling of 
the transition from solar to anti-solar rotation see \cite{43}.

Less evident options are sun-like stars substantially cooler than the Sun, very young stars and stars which 
demonstrate various pulsations and oscillations. The trend for superflare activity to increase with decreasing 
temperature was established and explained in \cite{13}  (see also \cite{44}). Rapidly rotating young stars may have
negative $\alpha$ because of strong twisting of rising magnetic loops by the Coriolis force \cite{45}, so the young 
Sun probably had superflares.
Differential rotation in young rapidly rotating stars was recently observed \cite{46,47}.  For slower rotating sun-
like F, G, and K stars, discovered during HK Project, differential rotation here has been  investigated in \cite{48}.

\section{Discussion and conclusions}

In this paper we suggest a scenario which allows us to understand how a stellar dynamo can provide a solar-like 
star with a magnetic field whose energy is substantially larger than that of the Sun. This in turn can explain 
why superflares of {\lower.4ex\hbox{$\;\buildrel >\over{\scriptstyle\sim}\;$}}$10^{35}$erg can occur on 
solar-type stars -- and some stars emanate them several times in a limited time interval -- while radionuclide 
data do not provide any evidences for solar flares of comparable energy over the past eleven millennia.

Our scenario originates from the idea of \cite{1}, and suggests that the sign of the dynamo drivers in the 
superflaring stars can be opposite to that of the sun on a regular basis rather than reverse occasionally as a result of rare 
fluctuations. Stellar dynamo simulations for  stars with  differential rotation dependent mainly on latitude show 
a change from oscillatory to steady dynamo
action with a strong increase in magnetic energy when the sign of the dynamo number is reversed. The sign 
reversal can be associated with anti-solar differential rotation or, perhaps, with change of sign of the 
$\alpha$-parameter compared to that of the Sun.

We present arguments that the deviations of stellar hydrodynamics from the solar one which can provide the 
intensive magnetic field can happen at least in some of the stars with superflares. The main straightforward 
reason for such a deviation can be the fact that some superflaring stars belong to binary systems. Observations 
favour anti-solar rotation in close binaries \cite{42}. Other examples of superflaring stars include giants, low mass 
main sequence dwarfs or young rapid rotators.

At this stage of research it looks too early to insist that the scenario suggested can explain all cases of 
stellar superflares, however it  can
explain at least some of them. Of course, the scenario explains only how to get high magnetic energy, while the 
detailed mechanism of flare production requires more than just dynamo studies. Some discussion concerning such 
processes can be found in, e.g., \cite{29,49}.



\phantomsection
\section*{Acknowledgments}

DS acknowledges financial support from RFBR under grant 15-02-01407.
IU's contribution was made in the framework of ReSoLVE Centre of Excellence (Academy of Finland, project no. 
272157). LLK is thankful to the Russian Foundation for Basic Research for the support (project 17-02-00016).
MMK is grateful to the RFBR (grant 15-02-06271) and the Leading Scientific Schools project 9570.2016.2 for 
financial support. We are grateful to referees for useful 
comments.

\phantomsection
\bibliographystyle{unsrt}

\end{document}